\documentclass[aps,pra,twocolumn,showpacs,amsmath,amsfonts,amssymb,floatfix]{revtex4}
\usepackage{graphicx}
\usepackage{bm}
\begin{document}
\title{Gravitational red-shift and deflection of slow light}
\author{J. Dressel, S. G. Rajeev, J. C. Howell, and A. N. Jordan}
\affiliation{Department of Physics and Astronomy, University of Rochester, Rochester, New York 14627, USA}
\date{\today}
\newcommand{\field}[1]{\mathbb{#1}}
\newcommand{\set}[1]{\mathcal{#1}}
\newcommand{\pd}[2]{\frac{\partial #1}{\partial #2}}
\newcommand{\pdd}[3]{\frac{\partial^2 #1}{\partial #3 \partial #2}}
\newcommand{\defas}[0]{\overset{def}{=}}
\def\be{\begin{equation}}
\def\ee{\end{equation}}
\def\bea{\begin{eqnarray}}
\def\eea{\end{eqnarray}}
\def\la{\langle}
\def\ra{\rangle}
\begin{abstract}
We explore the nature of the classical propagation of light through media with strong frequency-dependent dispersion in the presence of a gravitational field.  In the weak field limit, gravity causes a redshift of the optical frequency, which the slow-light medium converts into a spatially-varying index of refraction. This results in the bending of a light ray in the medium.  We further propose experimental techniques to amplify and detect the phenomenon using weak value measurements.  Independent heuristic and rigorous derivations of this effect are given.
\end{abstract}
\maketitle
We live in an age where we constantly push technological boundaries to explore the subtleties of the quantum domain and to challenge our understanding of the microscopic world.  Occasionally these explorations lead us full circle back to the macroscopic classical regime to prompt new and interesting questions about previously studied topics.  These revisitations only deepen our understanding of the world.

 New materials with strong frequency-dependent dispersion are also being explored that allow for remarkably slow group velocities in the propagation of light \cite{Milonni2004}.  These new developments have raised a new question: can we actually measure the effect of gravity on slow light using a table-top device?  The purpose of this paper is to describe how this could be done.

One of the first tests of the General Theory of Relativity \cite{Einstein1916} was the confirmation of the bending of light near the sun in 1919, showing that gravity does affect the propagation of light \cite{Dyson1920}.  However, it has been difficult historically to test these effects in any compact way in a laboratory.  Between the fact that gravity is a rather weak effect and the fact that light typically travels at large speeds, any measurement of a gravitational effect on the propagation has required a large propagation distance -- much larger than any typical laboratory.

For Earth-confined experiments, however, the dominant measurable effect of gravity on light is the gravitational red-shift of the frequency, and not the bending of its trajectory (see e.g. \cite{Hartle2003}).  According to this effect, a photon starting at an initial height $0$ and climbing to a final height $y$ will have its frequency red-shifted in a weak gravitational field,
\begin{equation}
\omega' = \omega (1 - g y/c^2).
\label{eq: Redshift}
\end{equation}
The shift in frequency can be viewed as a manifestation of the equivalence principle: the crests of sequential light waves spread further apart as the beam climbs.  Alternatively, the shift of the frequency can be considered to arise from conservation of energy as the photon climbs a potential well.  As we shall see later, the height-dependence of the frequency continues to be the dominant effect of gravity during slow light propagation that also results in a slight bending of the trajectory.

The shift in frequency due to gravitation may be small, but Pound and Rebka managed to measure and verify it to within 10\% of the predicted value in 1959 using a height difference of only $22.5$ meters \cite{Pound1959}.  Their experiment used gamma rays emitted by the nucleus of Iron-57 to achieve the sharp spectral lines needed to measure the shift.  The M\"ossbauer effect, stating that atoms in a lattice may emit radiation from their nuclei with almost no recoil since the entire lattice collectively recoils, allowed the gamma rays to be emitted with minimal Doppler broadening, improving the precision.  Further tests have since been done using greater height differences and techniques to measure the effect of gravitational redshift to a precision of about one part in $10^{4}$ \cite{Vessot1980}.  The effect has even been necessary to include in the GPS guidance system \cite{Hartle2003}.  To the best of our knowledge, the Pound-Rebka experiment holds the record for the shortest height difference to see a gravitational effect on light.

The aim of this paper is to combine the physics of slow-light materials with gravitational effects on the light.  We show from fundamental principles that a light ray is predicted to follow a parabolic trajectory through the medium (to leading order in the gravitational field strength).  The bending is predicted to arise from the gravitational redshift, which is amplified by the strongly dispersive slow-light medium.  For typical experimental setups at the time of writing, we predict the expected deflection to be on the order of an Angstrom.   We also propose an experimental set-up to detect this very small beam displacement.
Hence, if the experiment proposed in this paper is realized, then the effects of the gravitational red-shift would beat the previous height record by many orders of magnitude.  We note that this question has also been recently examined independently by Kumar \cite{Kumar2007}. However both our qualitative explanation, as well as our quantitative predictions of this effect are very different.

\section{Estimate and Intuition}
Later in this paper we will present a general analysis of the gravitational effect on light in a strongly dispersive medium from a General Relativity perspective.  However, the basic intuition of the physics can be understood quite simply.  A slow-light medium has an index of refraction that sharply varies linearly with frequency in a particular frequency range.  A beam of light will be slightly gravitationally red- or blue-shifted as it changes height in the medium.   The combination of these effects gives the index of refraction an effectively linear height dependence, which is amplified by the steep dispersion relation.  The height dependence of the index of refraction will then be translated into a spatial deflection of the light beam as the beam propagates.  Therefore, by measuring the spatial deflection of the beam, one is indirectly measuring the effects of the gravitational red-shift.

Relying on the above intuition, we can now estimate the size of the effect in a table-top experiment. Assuming that the deflection height, $y$, will be small, the index of refraction $n(\omega)$ may be expanded to linear order in $y$ as follows,
\begin{equation}
n(\omega, y) = n_0 +  \frac{dn}{d\omega}\frac{d\omega}{dy} y.
\end{equation}
The gravitational red-shift (\ref{eq: Redshift}) implies that $d\omega/dy = - \omega g/c^2$.  The classical theory of optics \cite{Born1959} defines the phase velocity $v_p$ and group velocity $v_g$ of a wave packet as
\begin{eqnarray}
v_p &\defas& \frac{\omega}{k} = \frac{c}{n}, \label{vp} \\
v_g &\defas& \frac{d\omega}{dk} = \frac{c}{n+ \omega \frac{dn}{d\omega}}, \label{vg}
\end{eqnarray}
and using the dispersion relation $k = \omega n/c$, we can rewrite $\omega \frac{dn}{d\omega} = c\left[ 1/v_g - 1/v_p \right]$.  Hence, the effective index of refraction may be written as
\begin{equation}
  n(\omega, y) = n_0 - \frac{g y}{c^2}\frac{c(v_p - v_g)}{v_g v_p}.
\end{equation}
Using the approximation that $v_p \approx c$, and $v_g \ll v_p$, we can rewrite this in the simplest form that still preserves the characteristic velocities:
\begin{equation}
  n(\omega, y) \approx n_0 - \frac{g y}{c^2}\frac{v_p}{v_g}.  \label{result}
\end{equation}

This index of refraction leads to a deflection of the beam, given by solving the geometric optics equation \cite{Born1959},
\begin{equation}
\frac{d}{ds} \left[n(\omega, y) \frac{d{\vec r}}{ds}\right] = \nabla n.
\end{equation}
This is done by considering a beam initially pointing in the horizontal $(x)$ direction and taking the $x$-component of the above equation to find the conservation law,
\begin{equation}
n(\omega, y) \cos \theta = const,
\end{equation}
where $\theta$ is the downward angle of the beam from the horizontal.  The trajectory may be found by considering
\begin{equation}
   \tan \theta  = -\frac{dy}{dx} = \sqrt{\frac{n(\omega, y)^2}{n(\omega, 0)^2} -1} \approx \sqrt{-\frac{2 g}{c^2}\frac{v_p}{v_g} y},
\end{equation}
where the expansion is to first order in $g v_p z/c^2 v_g$, and we have approximated $n_0 \approx 1$.
This simple differential equation can now be solved for the vertical deflection $\Delta y$, given a horizontal displacement $L$, yielding
\begin{equation}
\label{eq: Intuitive Displacement}
\Delta y \approx - \frac{ g L^2 v_p}{2 c^2 v_g}.
\end{equation}
Similarly, the angle $\theta$ of the final beam relative to the horizontal is given by,
\begin{equation}
\label{eq: Intuitive Angular Displacement}
\theta = \frac{ g L v_p}{c^2 v_g} = - 2 \frac{\Delta y}{L}.
\end{equation}
The more careful analysis later in this paper confirms that these expressions are indeed the first order approximation to the vertical drop.

 Making an estimate, we assume plausible lab values of $L = 0.8 m$, $v_g = 10^2 m/s$, and $v_p = 3\times 10^8 m/s$, so the vertical deflection is estimated to be $\Delta \approx 1$\AA, and the angular deflection is estimated to be $\theta \approx 0.2 nrad$.  By varying $L$, or $v_g$, the deflection can be adjusted by several orders of magnitude around this estimate.

\section{Experimental Tests}
The nontrivial nature of this proposed experiment requires careful attention.  The assumption that the index of refraction varies as a function of height restricts the number of systems that can be used to deflect the light.  For example, electromagnetically induced transparency \cite{Boller1991}, nonlinear magneto-optical rotations \cite{Barkov1988}, coherent population oscillations \cite{Bigelow2003} or Raman-Nath interference in liquid crystal light valves \cite{Residori2008} can probably all achieve the experimental products needed for an observable signal.  However, all of these systems use a strong pump beam to prepare a coherence or interference in the medium.  The pump beam is also experiencing the gravitational redshift.  For pump beams of nearly the same frequency as the probe, the redshift of the probe is then washed out, because the index of refraction established by the pump beam also shifts in the same direction as the probe beam.  Since $\partial_y \omega \propto \omega$, only when the pump beam is dramatically different in frequency will a pump-prepared system have the ability to detect the red-shift.  In addition to the problems associated with a shifting dispersion, pump beams with a nonuniform intensity profile (e.g., Gaussian beams) also cause some guiding effects due to index of refraction changes.  Some absolute frequency systems also have technical difficulties.  For example, hot atomic vapors have gravitational and thermal density gradients that would usually overwhelm the small deflection \cite{Camacho2006}.

One can envision certain classes of systems that have promise to measure the deflection.  One class of systems is one in which a pump beam and probe beam propagate in opposite directions.  This is effectively the same as having a pump beam blue shifted while the probe is red shifted.  The deflection would be twice as large, because the dispersion is shifting in the opposite direction in frequency space as the probe as a function of height.  Another class of systems is one which uses fixed frequency, narrow resonance, solid state, bulk materials.  If the inhomogeneous broadening is negligible, the probe frequency can be very close to the resonance frequency to achieve small group velocities.  These experimental parameters are demanding.  

In order to measure these small gravitational effects, we enlist quantum mechanics for help. Interference effects can measure length differences of much less than a wavelength.  Indeed, Manly and Page have proposed to essentially directly measure the change in the index of refraction (\ref{result}) by interfering a red-shift beam with a blue-shifted beam in a fiber-based Sagnac interferometer \cite{Manly2001}.  Such an approach can also be taken in a slow-light medium.  We describe below another quantum approach using another kind of quantum interference.  In order to measure such a small deflection of a light beam, we propose to extend recent advances in precision metrology by amplifying the deflection using quantum ``weak value'' measurements \cite{Aharonov1988}.  Weak values have the property that the post-selected expectation value of an operator can exceed the eigenvalue bounds of the operator \cite{Pryde2005,Williams2008}.  Recently, Hosten and Kwiat successfully used optical weak value measurements to detect a 1 \AA \ deflection in an optical beam \cite{Hosten2008}.  In the Hosten-Kwiat experiment, the weak value operator was the polarization of the light beam, which was entangled with the transverse position degree of freedom.    Most slow-light materials do not have a polarization-dependent index of refraction, so we propose to use instead a `which-path' operator of an optical Sagnac interferometer \cite{Born1959} as shown in Fig. 1.   A laser source is at the input of a Sagnac interferometer.  If no other optical element is in the interferometer and with ideal alignment, all of the light will exit the input port of the interferometer.  The optical path
length of both directions are identical in a Sagnac because both paths simply traverse the same route but in different directions.  The reason all of
the light exits only the input (or bright) port is due to a relative $\pi/2$
phase shift for each reflection versus transmission through the beam splitter, resulting in exactly destructive interference for the {\it dark port}.   Using a tunable birefringent element (a Soleil Babinet compensator, SBC), we can break the symmetry and cause a relative tunable phase between the two directions in the interferometer.  For this gedanken experiment, we assume the light entering the interferometer is horizontally polarized.  The light that is propagating in the counter clockwise direction is rotated to be vertical via a half wave plate oriented at 45$^\circ$ with respect to the horizontal polarization.  We align the SBC such that the vertically polarized light receives a relative phase shift $\phi$ compared with the horizontally polarized light which is propagating in the clockwise direction.  The light propagating in the clockwise direction is then made to be vertically polarized.  The two paths then interfere at the beam splitter but the counter clockwise light has picked up a tunable relative phase shift of $\phi$.
\begin{figure}[tb!]
\includegraphics[width=8cm]{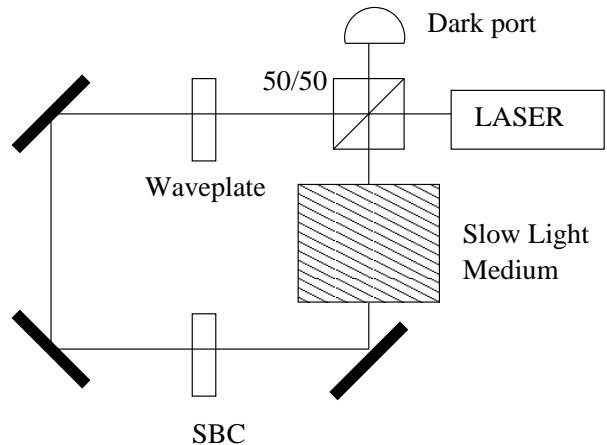}
\caption{Experimental Setup.  A laser beam enters a Sagnac
interferometer.  If no other optical element is in the system and
for perfect alignment all of the light exits out the input port of
the interferometer and goes back to the laser.  Using a
combination of a half ($\lambda/2$) wave-plate and a Soleil Babinet
compensator (SBC), one can tune the relative phase acquired for
each direction in the interferometer.  This allows us to tune
between the dark and bright ports, which is necessary for weak
value measurements.  The slow light medium the interferometer allows us to
couple the `which-way' information of the interferometer to the
transverse deflection of the light beam.  A quadrant detector, which measures
beam deflections, is in the dark port of the interferometer.}
\label{SpectralDeflection}
\end{figure}

A slow light medium (SLM) is placed in the interferometer, causing a
transverse gravitational deflection $\delta y$ of the beam in the downward direction.  We want to amplify this deflection using weak value techniques. By placing the SLM at an asymmetric point in the optical path, we can arrange that there is a long path of length $L_l$ (corresponding to the clockwise direction $\circlearrowright$ having a larger deflection $\delta y_l$) and a short path of length $L_s$ (corresponding to the counter-clockwise direction $\circlearrowleft$ having a smaller deflection $\delta y_s$) between the SLM and the 50/50 beam splitter.  We now describe how to use weak values to amplify the small spatial deviation.  We define the {\it system} as being the which-path information of the light beam, and the {\it meter}
as the transverse profile of the beam.  The system operator is the which-path operator ${\bf A} = \delta y_{l} |\circlearrowright\ra \la \circlearrowright| + \delta
y_{s} |\circlearrowleft \ra \la \circlearrowleft |$.  Following the propagation of the state through the Sagnac interferometer, starting from a pre-selected system state $| \psi_i\ra$ (formed at the BS), and postselecting on a system state $| \psi_f\ra$ (exiting the dark port) we find that the {\it weak value} 
\be 
A_w = \frac{\la
\psi_f | {\bf A} | \psi_i \ra}{\la \psi_f |  \psi_i \ra} 
\ee
of ${\bf A}$ is given by a purely imaginary result,
\be 
A_w \approx i \frac{\delta y_l - \delta y_s}{\phi}.
\ee 
The SBC phase $\phi \ll 1$ characterizes the deviation from perfect darkness, and is responsible for the amplification of the deflection.  See Ref.~\cite{us} for detailed calculations.  Expressing the deflection in terms of the gravitational deflection angle $\theta$ (\ref{eq: Intuitive Angular Displacement}), $\delta y_{l} -\delta y_s = (L_l - L_s) \theta$, we have for the total deflection,
\be 
\la \delta y\ra = \frac{F (L_l - L_s) \theta }{\phi},
\ee
where $F$ is a  factor originating from the dynamical propagation, which brings additional enhancement (a factor of about 100 in Ref.~\cite{Hosten2008}).
This deflection is then directly measured with a quadrant detector in the dark port.

\section{Hamilton's Optics}
In order to give a rigorous treatment of light propagation through media with anomalous dispersion in a gravitational field, we start with the wave equation itself.  Many of the historical treatments of light propagation have implicitly assumed the index of refraction to be weakly dependent (or not dependent) on frequency since the derivations were done before slow light was a commonly studied phenomenon, e.g. Ref.~\cite{Gordon1923}.  Hence, a revisitation of the physical foundations is illuminating.

The components of an vector EM wave propagating through a material medium obeys the modified wave equation,
\begin{equation}
  \label{eq:Scalar Wave}
  \left(-\frac{n^2}{c^2}\frac{d^2}{dt^2} + \nabla^2\right)\Psi = 0.
\end{equation}
Here $\Psi$ is any component of a vector EM wave, $n$ is the index of refraction of the material, and $c$ is the speed of light in vacuum.

We assume the propagating wave components have the form $\Psi = A\exp(i S/\lambda)$, where $S(\vec{r},t)$ is the \textit{eikonal} of the wave \cite{Born1959}. Then, to leading order in $1/\lambda$, we get an equation for the eikonal itself in the geometrical ray approximation,
\begin{equation}
  \label{eq:Eikonal Flat Space}
  -\frac{n^2}{c^2}\left(\frac{d}{dt}S\right)^2 + |\nabla S|^2 = 0.
\end{equation}
The eikonal relation holds for all components of the vector EM wave so the vector nature of the EM wave itself is unchanged during the propagation as a ray.  By taking the leading order in $1/\lambda$, we assume $\lambda$ is small compared to the length scales through which the wave propagates, such that the wave behaves effectively as a ray.  Note that the ray may still undergo small angular deflections without violating this approximation, which is a crucial point for the results in this paper.

In small regions of space and time, the eikonal can be expanded in a series to first order in space and time,
\begin{equation}
  \label{eq:Eikonal Expansion}
  S \approx S_0 + \vec{r}\cdot\nabla S + t\pd{S}{t},
\end{equation}
where we have incorporated the wavelength $\lambda$ into the eikonal for notational simplicity.  Comparing this approximation to the eikonal for a plane wave, $S = \vec{k}\cdot\vec{r} - \omega t + \alpha$, we define the frequency and wave vector of the wave for a \textit{local} region where the eikonal can be approximated as planar,
\begin{eqnarray}
  \vec{k} &\defas& \nabla S \label{eq:Wave Number Definition Flat Space},\\
  \omega &\defas& -\pd{S}{t} \label{eq:Frequency Definition Flat Space}.
\end{eqnarray}

We then rewrite the eikonal equation (\ref{eq:Eikonal Flat Space}) in terms of the local wave vector and frequency for each infinitesimal region of space and time,
\begin{equation}
  \label{eq: Dispersion Squared Flat Space}
  \vec{k}\cdot\vec{k} - \left(\frac{\omega}{c}\right)^2n^2(\omega,\vec{r}) = 0.
\end{equation}
Here the potential dependence of the index of refraction, $n$, on the local frequency and position is written explicitly as a reminder.

The eikonal equation can be rearranged and square-rooted, giving the usual dispersion relation for the medium in a localized region,
\begin{equation}
  \label{eq: Dispersion Flat Space}
  c|\vec{k}| = \omega n(\omega, \vec{r}).
\end{equation}
The signs are chosen to keep the frequency, $\omega$, positive.

An analogy can be made between the mechanics of material particles and the motion of wave packets that exposes a new set of canonical variables and allows a systematic derivation of the equations of motion for the wave packet.

From Eqns.~(\ref{eq:Wave Number Definition Flat Space}, \ref{eq:Frequency Definition Flat Space}), we see that the eikonal, $S$, plays the role of the action of the system, $\mathcal{S}$, with the wave vector, $\vec{k}$, corresponding to the momentum of the wave, $\vec{p}$, and the frequency, $\omega$, corresponding to the energy, $H$. In particular, as is well known from Quantum Mechanics, the energy and momenta of a wave packet composed of plane waves are directly proportional to the frequency and wave numbers, respectively.  The scaling constant in Quantum Mechanics is $\hbar$, implying that the classical action is just the scaled eikonal for the wave packet.

The explicit correspondence for the translation to Hamiltonian dynamics is
\begin{eqnarray}
  \mathcal{S} &\leftrightarrow& S \nonumber\\
  \vec{q} &\leftrightarrow& \vec{r} \nonumber\\
  \vec{p} &\leftrightarrow& \vec{k} \\
  t &\leftrightarrow& t \nonumber\\
  H &\leftrightarrow& \omega \nonumber.
\end{eqnarray}
Using this correspondence, either the eikonal equation (\ref{eq: Dispersion Squared Flat Space}) or the dispersion relation (\ref{eq: Dispersion Flat Space}) can be understood to represent Hamilton-Jacobi equations of the system in an implicit form \cite{LL1976}.  Solving each for $\omega$ would give the standard form of the Hamilton-Jacobi equation using these identified canonical coordinates.  Furthermore, $\omega$ will be a conserved quantity since the equations are independent of $t$.  For convenience in referring to these equations, we will denote the eikonal relation as $F$ and the dispersion relation as $G$:
\begin{eqnarray}
  F(\vec{k},\omega) &\defas& \vec{k}\cdot\vec{k} - \left(\frac{\omega}{c}\right)^2n^2(\omega,\vec{r}) = 0, \\
  G(\vec{k},\omega) &\defas& |\vec{k}| - \frac{\omega}{c} n(\omega, \vec{r}) = 0.  \label{G}
\end{eqnarray}

Using the conserved quantity $\omega$ as the Hamiltonian of the system, we can write down Hamilton's equations of motion directly in the form,
\begin{eqnarray}
  \frac{d\vec{k}}{dt} = \dot{\vec{k}} &=& -\nabla\omega, \label{eq: Hamilton's Eq - Wave Vector}\\
  \vec{v}_g \defas \frac{d\vec{r}}{dt} = \dot{\vec{r}} &=& \pd{\omega}{\vec{k}}. \label{eq: Hamilton's Eq - Wave Position}
\end{eqnarray}
It becomes clear at this point that $\vec{v}_g$ must be interpreted as the \textit{group velocity} of a wave packet that obeys the eikonal equation, and not the phase velocity of the carrier wave itself.  Hamilton's equations of motion along with the eikonal equation completely determine the motion of a \textit{wave packet} treated as a point-particle through a material medium along a geometrical ray.  The geometrical ray approximation made to derive the eikonal relation implicitly assumes wave packet behavior.

If the form of the index of refraction $n$ is known, then $F$ or $G$ may be solved for $\omega$, and Hamilton's equations can be solved for the trajectory of the packet directly.

\section{Influence of gravity}

In order to include the influence of gravity on the slow light beam, we recall that the main gravitational effect for earth-based experiments is the gravitational red-shift.  The following simple rule can be used to make a \textit{weak field} translation of the dispersion relation to include gravitational effects,
\begin{equation}
  \omega \to (1 - U)\omega,
  \label{eq: Frequency Redshift}
\end{equation}
where $U=-GM/c^2 r$ is the scaled Newtonian gravitational potential.  We give a general relativistic justification of this rule in the appendix.  It is the spatial dependence of $U$ in the shift of the frequency that will cause a bending of the trajectory of light.

The gravitational redshift is very small, $U \ll 1$, so we expand the dispersion relation (\ref{G}) after the replacement (\ref{eq: Frequency Redshift}) to first order in $U$,
\begin{eqnarray}
  G(\mathbf{k},\vec{r}) &\approx& |\vec{k}| - \frac{\omega n}{c} + U\frac{\omega}{c}\left[n + \omega \pd{n(\omega)}{\omega}\right], \nonumber \\
  &\approx& |\vec{k}| - \frac{\omega}{v_p}\left(1 - U\frac{v_p}{v_g}\right), \label{eq:Dispersion2} \\
  &=& 0, \nonumber
\end{eqnarray}
where we have used the standard definitions of $v_g$ and $v_p$ (\ref{vp},\ref{vg}).  We will see in the next section that the definition of $v_g$ is effectively unchanged.
This final simplification hides the simple frequency redshift that is occurring, but makes the form of the first order correction in terms of the quantity $U v_p/v_g$ particularly apparent. 

\subsection{Hamilton's Equations}
We proceed to solve Hamilton's equations of motion with the frequency $\omega$ playing the role of the Hamiltonian of the system, and the new dispersion equation (\ref{eq:Dispersion2}) acting as the Hamilton-Jacobi equation.

Hamilton's equation (\ref{eq: Hamilton's Eq - Wave Position}) gives the vector group velocity of the packet.  The frequency $\omega$ involves only the quantity $|\vec{k}|$, so $\pd{\omega}{\vec{k}}\propto\hat{k}$. Therefore, we now focus on the magnitude of the group velocity. Applying a frequency derivative to (\ref{eq:Dispersion2}) yields
\begin{eqnarray}
  \pd{|\vec{k}|}{\omega} &=& \frac{(1 - U)}{c}\left[n(\omega) + \omega\pd{n(\omega)}{\omega}\right] \nonumber \\
  && - 2U\frac{\omega}{c}\pd{n(\omega)}{\omega} - U\frac{\omega^2}{c}
  \frac{\partial^2 n(\omega)}{\partial \omega^2}. \label{dkdw}
\end{eqnarray}

Now we make the slow light approximation that $n$ is linear in $\omega$.  Slow light occurs for a frequency band where the index of refraction $n$ has a steep linear dependence on $\omega$ with positive slope.  Therefore we neglect the last (second-order derivative) term in Eq.~(\ref{dkdw}), so
\begin{eqnarray}
  \vec{v}_g = \frac{d\vec{r}}{dt} &=& \pd{\omega}{\vec{k}} = \left( \pd{|\vec{k}|}{\omega} \right)^{-1} \hat{k}, \nonumber \\
  &\approx& \frac{c}{(1 - U)n + \omega (1 - 3U) \pd{n}{\omega}} \hat{k}.
  \label{eq: Group Velocity Raw}
\end{eqnarray}
We see that the group velocity is modified very slightly from its traditional form by the presence of the gravitational field.  However, the packet will always travel in the same direction as the wave vector.

Hamilton's other equation, (\ref{eq: Hamilton's Eq - Wave Vector}), will determine how the wave vector itself changes over time and thus how the trajectory will curve.  Since the only position dependence in $\omega$ is radial due to the presence of $U$, the gradient reduces to a radial derivative.  We can implicitly find $\partial_r \omega$ by applying a spatial derivative to $G$,
\begin{eqnarray}
  \pd{ }{r}\Big[ c|k|  &=& (1 - U)\omega n - \omega^2 U\pd{n}{\omega} \Big],\nonumber \\
  0 &\approx& \pd{\omega}{r}\left( (1 - U)n  + (1 - 3U)\omega\pd{n}{\omega}\right) \nonumber \\
  && - \omega\pd{U}{r}\left(n + \omega\pd{n}{\omega}\right).
\end{eqnarray}
Solving for $\partial_r \omega$ gives
\begin{equation}
  \pd{\omega}{r} \approx \left[\frac{n + \omega\pd{n}{\omega}}{(1 - U)n + (1 - 3U)\omega \pd{n}{\omega}}\right]\omega\pd{U}{r}.
\end{equation}
Then from (\ref{eq: Hamilton's Eq - Wave Vector}) we find the change in wave vector to be
\begin{eqnarray}
  \frac{d\vec{k}}{dt} = \dot{\vec{k}} &=& - \nabla\omega = - \pd{\omega}{r} \hat{r}, \nonumber \\
  &\approx& - \left[\frac{n + \omega\pd{n}{\omega}}{(1 - U)n + (1 - 3U)\omega \pd{n}{\omega}}\right]\omega\pd{U}{r} \hat{r}.
  \label{eq: Wave Vector Change Pre}
\end{eqnarray}

Now we linearize the dimensionless gravitational potential $U$ to give a more useful form for a lab, and evaluate the derivative in (\ref{eq: Wave Vector Change Pre}),
\begin{equation}
  U = -\frac{MG}{c^2(R + r)} \approx \frac{-gR}{c^2} + r\frac{g}{c^2},
\end{equation}
so,
\begin{equation}
  \label{eq: Wave Vector Change Raw}
  \dot{\vec{k}} \approx - \left[\frac{n + \omega\pd{n}{\omega}}{(1 - U)n + (1 - 3U)\omega \pd{n}{\omega}}\right]\omega \frac{g}{c^2}\hat{r}.
\end{equation}
Thus we see a net deflection of the wave vector having to do with the presence of the local gravitational acceleration $g$.  Here $R$ is the radius of the Earth, $M$ is the mass of the earth, $G$ is Newton's gravitational constant, and $r$ is the height displacement from the initial starting position of the packet.  Notice that the scaling factor in (\ref{eq: Wave Vector Change Raw}) is the ratio of the new group velocity to the normal group velocity.

Noting the fact that $g/c^2 \approx 10^{-16}m^{-1}$, $g R/c^2 \approx 10^{-11}$, and $\omega \approx 10^{10} s^{-1}$ for the surface of the Earth and the optical domains we are considering, we can reduce equations (\ref{eq: Group Velocity Raw}) and (\ref{eq: Wave Vector Change Raw}) to simpler final forms,
\begin{eqnarray}
  \vec{v}_g = \dot{\vec{r}} &\approx& \frac{c}{n + \omega\pd{n}{\omega}}\hat{k}, \label{eq: Group Velocity}\\
  \dot{\vec{k}} &\approx& - \frac{g}{c^2}\omega \hat{r}. \label{eq: Wave Vector Change}
\end{eqnarray}
The group velocity appears unchanged to a very good approximation with such a weak field from what is traditionally expected in a slow light medium, but the wave vector picks up a small, effectively constant, deflection over time that is dependent on the carrier frequency of the packet and the local gravitational acceleration.  The deflection points toward the gravitating body, so light will bend toward the body, matching intuition.

\subsection{Wavepacket trajectory}
With the newly derived approximate equations of motion for the weak field limit at the surface of the Earth, (\ref{eq: Group Velocity}) and (\ref{eq: Wave Vector Change}), we can solve the trajectory of a point-like wave packet in a straight forward manner.

Assume that for a small local bit of trajectory the radial direction can be well approximated by the vertical Cartesian direction $\hat{y}$, and consider the wave vector pointing entirely in the horizontal Cartesian direction $\hat{x}$ at an initial time $t_i = 0$, so,
\begin{equation}
  \vec{k}(0) = \frac{\omega n}{c}\hat{x} = \frac{\omega}{v_p}\hat{x}.
\end{equation}
Here $v_p$ is the phase velocity of the packet.

Solving for $\vec{k}(t)$ from (\ref{eq: Wave Vector Change}) yields
\begin{eqnarray}
  \vec{k}(t) &=& -\frac{\omega g}{c^2} t \hat{y} + \vec{k}(0), \nonumber \\
  &=& \omega\left(\frac{1}{v_p}\hat{x} - \frac{g}{c^2} t \hat{y}\right), \label{eq: Wave Vector Time Dependent}
\end{eqnarray}
so
\begin{eqnarray}
  \hat{k}(t) &=& \frac{\vec{k}}{|k|}, \nonumber\\
  &=& \left[1 + \left(v_p \frac{g}{c^2} t\right)^2\right]^{-1/2}\left(\hat{x} - v_p \frac{g}{c^2} t \hat{y}\right), \nonumber \\
  &\approx& \hat{x} - v_p \frac{g}{c^2} t \hat{y}  \label{approx}.
\end{eqnarray}
Thus the wave vector rotates over time, and its magnitude increases due to the  gravitational redshift.  Notice that the wave rotation is not dependent on the group velocity, but only on the carrier phase velocity.  The group velocity only appears when we calculate the position of the packet.  The final approximation (\ref{approx}) is made since we are only concerned with the lowest-order drop \cite{note1}. 

Solving for $\vec{r}(t)$ from (\ref{eq: Group Velocity}), starting the packet at an initial position $\vec{r}(0)$ at the start of a medium of length $L$ yields
\begin{eqnarray}
  \vec{r}(t) &=& v_g \int \hat{k}(t) dt, \nonumber \\
  &\approx& \vec{r}(0) + v_g t \hat{x} - \frac{1}{2}\frac{g}{c^2}v_p v_g t^2 \hat{y},
  \label{eq: Time-Dependent Trajectory Approximation}
\end{eqnarray}
which is a parabolic trajectory analogous to a kinematic trajectory, but with an effective gravitational acceleration given by,
\begin{equation}
  g_{\rm eff} \defas g \frac{v_p v_g}{c^2}. \label{eq: Effective Gravitational Acceleration}
\end{equation}
Thus a wave packet drops much {\it less} due to gravity in a medium with slow group velocity than it would in vacuum in the same amount of time, explaining why we have not casually observed them dropping like rocks in previous experiments!

Since we are interested mostly in how the wave packet trajectory is altered after traversal though a finite slow-light region, we set the final horizontal position to the length of the propagation region: $r_x(t_f) = L$.  Solving for $t_f$, we see that
\begin{equation}
  t_f = \frac{L}{v_g}.
\end{equation}

We can now solve for the vertical drop after a traversal of horizontal distance $L$, yielding,
\begin{eqnarray}
  \Delta y &=& - \frac{g}{c^2}\frac{v_p}{v_g}\frac{L^2}{2}.
  \label{eq: Vertical Displacement Approximation}
\end{eqnarray}
Again, this drop is analogous to the drop found in a kinematic trajectory, but with the effective gravitational acceleration (\ref{eq: Effective Gravitational Acceleration}) and horizontal velocity $v_g$.  Despite the fact that the effective gravitational acceleration is reduced, the long propagation time compensates for this, resulting in an amplified deflection. 

\section{Conclusion}
We have shown that a slow light medium is expected to amplify the effects of gravity on a wave packet such that they may become visible to a sensitive apparatus using weak value measurement.  The primary weak field gravitational effect is the induced position-dependence of the carrier frequency of the light, $\omega \rightarrow (1 - U)\omega$, where $U$ is the dimensionless Newtonian gravitational potential.  The position-dependence of the frequency shift is amplified by the strong frequency dependence of the slow light medium, which translates to a magnified bending of the trajectory of the light as it moves through the medium.

Using the modified dispersion relation for the EM wave components of the light (\ref{eq:Dispersion2}) as a Hamilton-Jacobi equation, we applied Hamilton's equations of motion to find a parabolic trajectory of the slow-light pulse through the medium  (\ref{eq: Time-Dependent Trajectory Approximation}).
We found that the trajectory  is analogous to a simple kinematic trajectory, but with a much smaller effective gravitational acceleration.  The smaller effective acceleration is more than compensated by the longer travel time, leading to the amplified vertical drop (\ref{eq: Vertical Displacement Approximation}).
For typical current lab values, we expect a vertical drop on the order of $1$ Angstrom for a traversal distance of roughly $1 m$.  

Although the predicted displacement is tiny, it borders what is currently possible to detect with the amplified precision of a weak value enhanced measurement.  We have proposed an optical weak values measurement, based on post-selected weak measurements of a which-path operator in a Sagnac interferometer.  This proposal has the advantage that in contract to previous experiments \cite{Hosten2008}, the optical deflection does not need to be polarization dependent.  We stress that this fact makes the idea applicable to amplify an optical deflection originating from {\it any} optical element.  Aside from the fundamental aspects of detecting a gravitational red-shift in a laboratory setting, a table-top detector would open a new field in precision gravitational metrology.

\appendix
\section{}
The purpose of this appendix is to describe how to relativize slow-light equations of motions.  We will convert the Eikonal equation into relativistic form, then pass to curved space-time, and then specialize to the weak-field case in order to justify the red-shift replacement $\omega \rightarrow (1-U) \omega$.  We will also comment on other methods that we have used to independently check the results in the paper.

{\it Flat Spacetime.}---
The equations developed for Euclidean flat space parametrized by a universal time can be generalized to a Minkowski spacetime in a straight-forward manner.  The generalization involves moving from the flat product space $\field{R} \times \field{R}^3$ with Euclidean metric on each piece to the flat spacetime manifold $\set{M}$ with a pseudometric having signature $(-,+,+,+)$.

To convert the flat space eikonal relation (\ref{eq: Dispersion Squared Flat Space}) into the language of spacetime, we note that the eikonal is a scalar field on the manifold.  The natural replacement for the frequency and wave number of the wave is the oneform field (written in boldface) defined as the exterior derivative of the eikonal,
\begin{equation}
  \label{eq: Wave Number}
  \mathbf{k} \defas dS = \pd{S}{x^\mu}dx^\mu = k_\mu dx^\mu.
\end{equation}
We use Einstein summation convention so all doubled Greek indices have implied sums over all 4 dimensions of spacetime, and Latin indices have implied sums over only the 3 spatial dimensions of spacetime.  

Written in the component representation with the coordinate basis the wave number oneform field is
\begin{equation}
  \mathbf{k} \doteq \left(\pd{S}{(ct)}, \pd{S}{x}, \pd{S}{y}, \pd{S}{z}\right) = \left(-\frac{\omega}{c},k_x,k_y,k_z\right).  \label{kvec_def}
\end{equation}
Here the notation $\doteq$ indicates an isomorphism to $\field{R}^4$ through a particular component representation for convenience.  In order to properly translate the eikonal equation (\ref{eq: Dispersion Squared Flat Space}) to spacetime, we will treat the appropriate components as the components of this oneform field.

Keeping in mind that the dot product in the eikonal equation (\ref{eq: Dispersion Squared Flat Space}) should correspond to application of the metric, we get the following translation into spacetime, in component form,
\begin{equation}
  \label{eq: Dispersion Squared Spacetime}
  F(\mathbf{k},\vec{r}) \defas \eta^{ij}k_ik_j + \eta^{00}k_0k_0n^2(k_0, \vec{r}) = 0.
\end{equation}

Notice that this equation is not Lorentz invariant unless $n = 1$.  The presence of the medium breaks the symmetry and forces a preferred frame for interpretation of the components of the wave vector.  This is precisely why the components have to be interpreted in the correct frame for the lab measurement.  Assuming the lab is stationary, the correct frame is the coordinate frame.  Changing to different frames results in frequency contraction (Doppler) effects from the lack of Lorentz invariance.

{\it Curved Spacetime.}---
The transition to curved spacetime with metric $g$ follows by considering the spacetime version of the eikonal equation (\ref{eq: Dispersion Squared Spacetime}) to remain valid for any orthonormal frame and co-frame \cite{Gockler1990}.  It is always possible in a curved spacetime to find an orthonormal frame and co-frame, though they will \textit{not} be a coordinate frame pair.

The key assertion is that the spacetime eikonal equation (\ref{eq: Dispersion Squared Spacetime}) is valid in this new orthonormal co-frame ($k_\mu \rightarrow {\tilde k}_\mu$).  Finding the proper form of the equation for the lab frame requires performing a frame change. It is important to interpret the components of the one form field $\mathbf{k}$ as a measurable frequency and wave number only in the correct frame--namely the lab frame--due to the fact that the components will change with the frame.  Note in particular that there is no coordinate change happening during this process.  Everything on the manifold will still be parametrized by the original Cartesian coordinates.

The frame-change results in a frame-transformation matrix, $e$, that is related to the Minkowski metric, $\eta$ , the metric of curved space, $g$, and the oneform field, $\mathbf{k}$, as,
\begin{eqnarray}
  \tilde{k}_\mu &=& e_\mu^\nu k_\nu, \\
  \tilde{k}^\mu &=& (e^{-1})_\nu^\mu k^\nu, \\
  \eta_{\mu\nu} &=& e^\alpha_\mu e^\gamma_\nu g_{\alpha\gamma}, \\
  \eta^{\mu\nu} &=& (e^{-1})_\alpha^\mu  (e^{-1})_\gamma^\nu g^{\alpha\gamma}.
\end{eqnarray}
If the form of the metric is known for a particular lab co-frame (usually the coordinate co-frame), then the form of the transformation matrix $e$ is completely specified \cite{Hartle2003,Gockler1990}.

Using these the above relations to transform the eikonal equation to the coordinate co-frame, we find that,
\begin{eqnarray}
  F(\mathbf{k},\vec{r}) &\defas& \eta^{ij}\tilde{k}_i\tilde{k}_j + \eta^{00}\tilde{k}_0\tilde{k}_0n^2(\tilde{k}_0, \vec{r}), \nonumber \\
 &=& g^{\alpha\gamma} k_\nu k_\mu \big[\left((e^{-1})_\alpha^i e^\nu_i\right) \left((e^{-1})_\gamma^j e^\mu_j\right) + \nonumber\\
 &&\left((e^{-1})_\alpha^0 e^\nu_0\right) \left((e^{-1})_\gamma^0 e^\mu_0\right)  n^2( e^\rho_0 k_\rho, \vec{r}) \big], \nonumber \\
 &=& 0. \label{eq: Dispersion Squared Curved Spacetime}
\end{eqnarray}
The wave oneform components in the lab frame are now precisely in the form we can interpret physically as in Eq.~(\ref{kvec_def}).
Note that the matrices from the co-frame changes do not completely cancel due to the lack of Lorentz invariance in the equation.  Furthermore, the index of refraction gains a spatially-dependent part.  All the components of the wave oneform become mixed in the equation on frame changes, leading to various frequency contraction (Doppler) effects in different frames.

In the special case when $n = 1$, the co-frame change matrices do exactly cancel and the eikonal equation becomes Lorentz invariant.  This special case is the traditional case for vacuum propagation, $F_{n=1} = k_\mu k^\mu =0$.

The new eikonal equation (\ref{eq: Dispersion Squared Curved Spacetime}) gives us a Hamilton-Jacobi equation for the propagation of a wave packet through \textit{curved} spacetime in the geometrical optics approximation in the lab frame, assuming the medium through which it propagates has no other effect than to introduce the index of refraction into the wave equation.  In particular, any microscopic interaction in the medium which would prevent the packet from falling due to gravity is being ignored.

As before, if $n$ is independent of $t$, then the entire eikonal equation is independent of $t$.  Therefore, $\omega$ becomes a conserved scalar quantity of the motion.  It is known as the ``world frequency'' since it is invariant with respect to the world time.  The ``proper frequency'' is conjugate to the proper time of an observer, rather than the world time, and differs by a frame transformation matrix from the coordinate frame to the observer frame.

{\it Weak Field Limit.}---
In order to compute something measurable, we use the weak-field limit of the Schwartzchild metric for a radially symmetric gravitating body (the Earth).  In Cartesian coordinates the non-flat part of the weak-field metric and the transformation matrix from flat spacetime are given by the relations,
\begin{eqnarray}
  r_s &\defas& \frac{2GM}{c^2}, \\
  U &\defas& -\frac{r_s}{2r}, \\
  g^{00} &\approx& -(1 - U)^2 \approx -(1 - 2U), \\
  g_{00} &\approx& -(1 + U)^2 \approx -(1 + 2U), \\
  e^{-1} &\approx& \mbox{diag}\left( 1 + U, 1, 1, 1 \right), \\
  e &\approx& \mbox{diag}\left( 1 - U, 1, 1, 1 \right),
\end{eqnarray}
where $r_s$ is the Schwartzchild radius of the gravitating body, and $U$ is the dimensionless Newtonian gravitational potential.  The rest of the metric $g$ is identical to the flat space metric $\eta$ in this weak-field limit.

The eikonal equation (\ref{eq: Dispersion Squared Curved Spacetime}) then reduces to the form:
\begin{eqnarray}
  F(\mathbf{k},\vec{r}) &\defas& g^{\alpha\gamma} k_\nu k_\mu \big[\left((e^{-1})_\alpha^i e^\nu_i\right) \left((e^{-1})_\gamma^j e^\mu_j\right) + \nonumber \\
 && \left((e^{-1})_\alpha^0 e^\nu_0\right) \left((e^{-1})_\gamma^0 e^\mu_0\right)  n^2( e^\rho_0 k_\rho, \vec{r} ) \big] \nonumber \\
 &=& g^{ij} k_i k_j +  g^{00} k_0 k_0 n^2(e_0^0 k_0, \vec{r}) \nonumber \\
 &=& \vec{k}\cdot\vec{k} - (1 - 2U)\left(-\frac{\omega}{c}\right)^2 n^2\left( -(1 - U)\frac{\omega}{c}, \vec{r}\right) \nonumber \\
 &=& \vec{k}\cdot\vec{k} - \left(\frac{(1 - U)\omega}{c}\right)^2 n^2\left( (1 - U)\omega, \vec{r}\right) \label{eq: Dispersion Squared Weak Field}\\
 &=& 0 \nonumber
\end{eqnarray}
The last step simply omits constants in the functional form of $n$, showing the relevant dependence, and puts the gravitational dependence in a more consistent form.  Notice that the simple diagonal form neatly keeps the frequency components unmixed, and only leaves gravitational influence attached to $\omega$ as a small correction.  Since the dimensionless potential $U$ is dependent purely on the radius, this equation is invariant in form with respect to a coordinate change to polar coordinates as well.
Taking the square-root of this equation gives the modified dispersion relation, justifying the gravitational red-shift replacement (\ref{eq: Frequency Redshift}).

We now briefly describe other independent checks done on our results.
Rather than solve Hamilton's equations in three dimensions, it is also possible to give a four dimensional derivation.  In the 4D approach the frequency is conserved, and that quantity $F$ or the quantity $G$ are treated as the Hamiltonian and an arbitrary global parameter $\tau$ is introduced.  The equations of motion eventually reduce to the 3D ones \cite{jd}.

\end{document}